# Strain-driven non-collinear magnetic ordering in orthorhombic epitaxial YMnO$_3$ thin films

X. Martí[1], V. Skumryev[2,3], V. Laukhin[1,2], R. Bachelet[1], C. Ferrater[4], M.V. García-Cuenca[4], M. Varela[4], F. Sánchez[1], J. Fontcuberta[1]

[1] Institut de Ciència de Materials de Barcelona (ICMAB-CSIC), Campus UAB, 08193 Bellaterra, Spain

[2] Institució Català de Recerca i Estudis Avançats, Lluís Companys 23, 08010 Barcelona, Spain

[3] Universitat Autònoma de Barcelona, Dept. Física, Campus UAB, 08193 Bellaterra, Spain

[4] Universitat de Barcelona, Dept. Física Aplicada i Òptica, Martí i Franquès 1, 08028, Barcelona, Spain

We show that using epitaxial strain and chemical pressure in orthorhombic YMnO$_3$ and Co-substituted (YMn$_{0.95}$Co$_{0.05}$O$_3$) thin films, a ferromagnetic response can be gradually introduced and tuned. These results, together with the measured anisotropy of the magnetic response, indicate that the unexpected observation of ferromagnetism in orthorhombic o-RMnO$_3$ (R= Y, Ho, Tb, etc) films originates from strain-driven breaking of the fully compensated magnetic ordering by pushing magnetic moments away from the antiferromagnetic [010] axis. We show that the resulting canting angle and the subsequent ferromagnetic response, gradually increase (up to ~ 1.2º) by compression of the unit cell. We will discuss the relevance of these findings, in connection to the magnetoelectric response of orthorhombic manganites.

# I. INTRODUCTION

Multiferroic materials are being intensively investigated due to their potential application in spintronics. Particularly relevant is the recent discovery that pronounced dielectric anomalies or even ferroelectricity can be obtained, in the so-called improper ferroelectrics, subsequent to magnetic ordering (see for instance [1]). In this case, either particular non-collinear magnetic structures (such as *spiral* order) or even collinear ordering can lead to ferroelectricity. The best known examples for this class of materials are probably the antiferromagnetic orthorhombic manganites (o-RMnO$_3$) where, depending on the Mn-O-Mn bond angle, different magnetic structures can be obtained ranging from collinear (E-type AFM) such as observed in TmMnO$_3$ [2] or HoMnO$_3$ [3] to sinusoidal or spiral as in YMnO$_3$ [4] or TbMnO$_3$ [5] respectively. Ferroelectricity has been observed in all of them. Apart from these, let's call them intrinsic ferroelectrics, some particular type of domain walls, as recently emphasized by D. Komskii [1], can also be the source of electric polarization and, in antiferromagnets, lead to magnetic frustration and, eventually, to some ferromagnetic response. This novel approach may have profound implications as it may constitute a new way to obtain improper ferroelectric and, if domain walls in antiferromagnets were a source of incomplete magnetic compensation, ferromagnetism and ferroelectricity would be intimately linked.

From these considerations it follows that epitaxial thin films of antiferromagnetic o-RMnO$_3$ oxides offer unique opportunities to tailor the magnetic structure eventually breaking antiferromagnetic ordering and lowering its symmetry or even engineering crystal domains and/or magnetic domain walls. Indeed, recent results in o-RMnO$_3$ thin films [6-13] brought the unexpected observation of spontaneous magnetization, in the following labelled as weak-ferromagnetism (w-FM), in what it was supposed to be antiferromagnetic compounds. However, the understanding and control of this w-FM has remained elusive: whereas earlier reports suggested strain-driven effects [6-9], it has also been claimed that it could be originated from symmetry breaking at domain walls [13]. The observation of w-FM in these materials may point to the occurrence or modification of an eventual of magnetoelectric coupling. However, the lack of a detailed understanding of the observed w-FM may have hampered its exploitation to induce magnetoelectric response.

Here we will show that by an appropriate use of epitaxial strain, the magnetic properties of o-YMnO$_3$ can be gradually tuned into a spin-canted away from the b-axis magnetic structure. In this manuscript we follow different strategies to allow fine tuning of structural parameters and structural strain: (i) film-thickness variation, (ii, iii) growth and annealing conditions and (iv) chemical composition (Co substitution). As 3d$^6$-Co$^{3+}$ ions have the same magnetic moment as 3d$^4$-Mn$^{3+}$ cations but smaller ionic radius, partial substitution of Mn by Co is an additional fine-tuning tool for the shrinking of unit cell and of the magnetic interactions. It will be shown that the (ferro)magnetic response, that is the occurrence of a finite net magnetic moment, does not correlate with the film roughness thus denying spin incomplete magnetic compensation at grain boundaries as a source of w-FM. Detailed data analysis, particularly the observed magnetic anisotropy, indicates that ferromagnetic response arises from strain-unbalanced antiferromagnetic structure. Therefore, it follows that epitaxial strain can be used to induce and tailor the ferromagnetic response in otherwise antiferromagnetic materials, thus opening a way to obtain magnetic structures susceptible of tunable magnetoelectric coupling. In fact, in a subset of the samples presented here, indications of magnetoelectric coupling have been already reported [14]. Our findings, which specifically refer to o-YMnO$_3$, are not, however, restricted to this oxide as other oxide thin films of the same o-RMnO$_3$ family are expected to display a similar response.

## II. EXPERIMENTAL

Thin films were deposited on [001] and [110]-oriented SrTiO$_3$ (STO) substrates by pulsed laser deposition. A KrF excimer laser beam (repetition rate of 5 Hz) was focused on stoichiometric YMnO$_3$ (YMO) and YMn$_{0.95}$Co$_{0.05}$O$_3$ (YMCO) targets placed at a distance of 5 cm of the substrate. Samples were grown at 785 ºC and P$_{O2}$ in the 0.1 – 0.3 mbar range. By changing the number of laser pulses, samples with different thickness (30 – 140 nm) were obtained. Details on YMO samples can be found elsewhere [8,15]. Film thickness (t) was determined by X-ray reflectivity. X-ray diffraction (XRD) measurements were carried out with a Philips' Material Research Diffractometer. Magnetic data were obtained using a Superconducting Quantum Interference Device by Quantum Design. Atomic force microscopy images were obtained with PicoSPM from Molecular Imaging.

## III. RESULTS

**A. Tuning the lattice parameters**

Fig. 1(a) shows the θ/2θ XRD scan for one illustrative YMCO sample, showing only (00l) peaks (Pbnm setting of YMO) with no traces of other phases and/or orientations. φ-scan (Fig. 1(b)) reveals that the film is epitaxial. The splitting observed in the zoom of φ-scan (Fig. 1(c)) indicates the presence of two in-plane crystal domains, 90 degrees in-plane rotated, with the epitaxial relationships [100]YMCO//[110]STO and [010]YMCO//[110]STO. The same crystal domain structure was observed in all YMCO and YMO films [15].

We previously reported [8] that lattice parameters and cell volume of YMO films change progressively with oxygen pressure during growth or increasing thickness. Here we will show that lattice parameters of YMO can be changed also by Co-doping. The reciprocal space maps around the (208) and (028) reflection for two pairs of YMO and YMCO samples having the roughly same thicknesses (~ 140 nm) but grown under distinct $P_{O2}$ (0.3 and 0.1 mbar) are shown in Fig. 2 (panels a and b, respectively). Lines in Fig. 2 indicate the corresponding position of bulk (208) and (028) reflection of YMO [16]. We discuss first data corresponding to films prepared at the highest $P_{O2}$ (Fig. 2(a)). Regarding the in-plane H directions, it turns out that, in both compounds, the (208) reflections appear at the bulk position, thus indicating that the cell along *a* is fully relaxed. In contrast, when scanning along the in-plane K direction, it appears that the (028) reflections are shifted towards higher K values. It indicates that both YMO and YMCO films have smaller *b* parameters than bulk YMO; therefore their unit cells are compressively strained along *b*. Regarding the out-of-plane direction L, data in Fig. 2(a) indicate that the (208) and (028) reflections of both YMO and YMCO appear at smaller L-values than bulk, thus indicating an expansion of the *c* parameter. In summary, the epitaxial strain is anisotropic being tensile along [001], compressive along [010] and negligible along [100]. By comparing data in Fig. 2 for YMCO and YMO films, it is observed that shorter *b* and larger *c* and are obtained by partial Co-substitution. As a result of the changes in the lattice parameters, the unit cell volume $V_c$(YMCO) is smaller than that of $V_c$(YMO). For instance, for samples in Fig. 2(a), $V_c$(YMCO) = 223.19 Å$^3$ and $V_c$(YMO) = 224.37 Å$^3$, both visibly smaller (compressed) than the YMO bulk value (226.47 Å$^3$ [16]). Inspection of data of YMO and YMCO films prepared at 0.1 mbar (Fig 2(b)) indicate the very same trend although obviously the cell parameters are different. Therefore, varying film thickness [8], growth conditions ($P_{O2}$) and Co-substitution, the cell parameters and unit cell volume can be varied at whish.

Post-growth annealing of the thin films is an alternative method to modify the unit cell parameters. Annealings effects under air or Ar atmosphere were tested on a YMO film relatively thin (30 nm) grown at 0.1 mbar $P_{O2}$. A first annealing step (Anneal 1) was performed on air during 12 h at 800 ºC; in a second annealing step (Anneal 2) the very same sample was annealed again for 2 h but in Ar atmosphere (0.1 mbar) at 700 ºC. The reciprocal space maps around the (208) and (028) YMO reflections were recorded after each annealing step and are shown in the left and right columns of Fig. 3, respectively. The bulk positions for the displayed diffraction spots are signalled by vertical and horizontal lines. The vertical dashed line is a guide for the eye. Data show that the *a*- and *b*-cell-parameter (H and K-directions) do not change appreciably by air annealing (Anneal 1) and remain closely coincident with those of the as-grown YMO film (top panels in Fig. 3) and bulk YMO. In contrast, the *c* cell parameter slightly shortens and approaches the bulk value; in overall the unit cell is somewhat compressed (0.05 %). Both oxygen uptake and plastic relaxation may contribute to the observed behaviour. A later reducing (Ar atmosphere) annealing process (Anneal 2) of the same sample promotes an expansion of the *b*-axis whereas the others (*a* and *c*) remain virtually constant and thus some expansion of the unit cell (0.72 %) occurred suggesting that desorption of oxygen may have occurred leading to an expansion of the unit cell volume. In short, film annealing under distinct atmospheres provides an additional tuning tool to modulate the cell parameters of YMO perovskite.

**B. Magnetic properties versus lattice parameters**

We now focus on the magnetic properties of the YMCO films. Detailed results on strained YMO films can be found elsewhere [8]. Field-cooled (FC) and zero-field cooled (ZFC) magnetization measurements have been performed in the 150 K to 5 K temperature range with 500 Oe magnetic field applied in-plane of the sample. In Fig. 4(a) we include the FC data for selected samples (samples are labelled by the corresponding $V_c$; the corresponding thickness (t) and used $P_{O2}$ values, indicated in the caption). The ZFC branches are shown, for sake of clarity, only for two samples. Thermal ZFC-FC hysteresis is well evident at low temperature revealing the occurrence of net magnetization below the antiferromagnetic ordering temperature signalled by a kink well visible at $T_N \approx 40$ K. Data in Fig. 4(a) clearly shows that the low-temperature

FC magnetization, is much enhanced as $V_c$ shrinks. This behaviour resembles the trend we reported for the pure YMO films [8] where a correlation of net ferromagnetic moment with the strain of the *b* parameter was found. To further determine if the distortion along the modulation direction of the antiferromagnetic structure (*b*-axis) is the only relevant parameter accounting for the observed response, we compare in Fig. 5(a) the FC magnetization at 25 K (measured at 500 Oe) versus *b*- and *c*- lattice parameters; *a* is not discussed because, as shown above, for YMCO films, and in agreement with previous results for YMO [8], it remains constant. Data in Fig. 5(a) shows that for both *c*- and *b*- parameters (top and bottom panels respectively), a monotonic increase of the magnetization is observed as *b* shrinks or *c* increases, indicating that the more distorted the unit cell is, the stronger is the (ferro)magnetic response. However, comparison of data for YMO and YMCO in Fig. 5(a) (bottom) display distinct slopes in (M(25K, 500Oe) *vs b*) thus indicating that the observed w-FM is not solely determined by the length of the *b*-axis. Most naturally, the contribution of the modification of *c* (expansion) and *a* (constant) axis upon epitaxial and chemical strain should be also included to get a measure of the overall distortion of the unit cell. Note that the ground magnetic structures in bulk o-$RMnO_3$ are essentially determined by the Mn-O-Mn bonding angle and subsequent modification of the strength of the competing magnetic interactions which should be modified through changes in both *b* and *c* lattice parameters. Therefore, we chose the unit cell volume as the simplest parameter to include all structural distortions and we plot in Fig. 5(b) M(25 K, 500Oe) (accounting for net magnetic moment) versus $V_c$ (accounting for unit cell distortion). Collapsing of all data (corresponding to YMO, YMCO films of different thicknesses and growth conditions) onto a single curve strongly indicates that the overall distortion of the unit cell is determining the induced ferromagnetism rather than the simple shrinking of the *b*-axis. Therefore we conclude that the observed w-FM results from unbalanced magnetic Mn-O-Mn(Co) interactions by modification of bond length topology along *b* and *c* due to epitaxial strain or chemical pressure.

It is worth mentioning that since unit cell volume for all films is slightly smaller than the bulk, the eventual presence of oxygen vacancies are not expected to play a major role. The annealing experiments discard absolutely oxygen vacancies as cause of the observed w-FM. FC magnetization curves ($H_{//}$ = 500 Oe) measured for an YMO sample before and after the annealing described in the previous section are given in Fig. 4(b). Data show that, upon the Anneal 1 process (air, 800 ºC), the magnetization slightly reduces as a result of the tiny changes in the lattice parameters. More noticeable is the drastic reduction of susceptibility and thus of w-FM after Anneal 2 process (Ar, 700 ºC, favouring the creation of oxygen vacancies). In correspondence to the visible expansion of the unit cell (Fig. 3), there is drop of over 60 % of the magnetization. Importantly, the magnetic and structural data of these annealed films fall just on top of the data shown in Fig. 5(b) for all films. We must therefore conclude that irrespective of the method used to tune the lattice parameters (thickness and $P_{O2}$ [8], chemical pressure or annealing) the ferromagnetic response is ruled by the unit cell distortion.

**C. Magnetic properties versus surface morphology**

Next, aiming to discriminate among strain and surface or domain boundaries as a source of w-FM by spin disorder, as suggested in Ref. 16, we compare the magnetic properties of films having distinct morphologies. Topographic images of representative YMCO and YMO films are shown in Figs. 6(a) and 6(b) respectively. Magnetometry revealed that the magnetization of the YMCO sample (0.84 emu/cm$^3$) is significantly smaller than in the YMO sample (3.11 emu/cm$^3$). Both images display morphology of a dense granular system with a significant quantitative difference: YMCO films are formed by smaller grains and thus the grain density and the density of grain boundaries are definitely larger than in YMO. If disorder at grain boundaries or surfaces were relevant sources of net magnetization, one would expect M to increase with grain density and roughness. The dependencies of the measured magnetization M(a25 K) on grain density and roughness (root mean square, RMS) for all samples shown in Fig. 6(c) and 6(d) clearly indicate the opposite trend, thus denying any important role of surface or interface disorder as sources of w-FM.

**D. Magnetic anisotropy**

Finally, we focus on the anisotropy of the magnetic response. In the simplest antiferromagnetic picture, a drop of magnetization below $T_N$ should occur if the measurement is performed with **H** applied along the antiferromagnetic axis whereas magnetization should remain constant if **H** is applied perpendicular to this axis. In YMO, according to neutron diffraction experiments [4], the spins are confined along the [010] direction with a sinusoidal modulation of their amplitude. We recall here that our films contain two-type of in-plane domains with (*a*, *b*) axis rotated in-plane by 90º. In this situation, when measuring in the plane of the sample along the two edges of the substrate ([100]$_{STO}$), **H** will be applied along the [110] direction of YMO, that is 45º away from the spin modulation axis (*b*). On the other hand, in the out-of-plane measurements, **H** is applied along the [001]$_{YMO}$ direction which is perpendicular to the *b* direction. Figs. 7(a) and 7(b) display the ZFC-FC curves for the most strained YMO sample we produced [8]. Notice that here magnetic moment rather than magnetization is displayed to emphasize that the features described below are a genuine experimental observation, not affected mass or volume normalization. Panels (a) and (b) correspond to the in-plane and out-of-plane measurements, respectively, and are sketched in Fig. 7(e). Data show that in both measurements the ZFC-FC curves split at around T$_N$ ~ 40 K indicating that the magnetic ordered state is not purely antiferromagnetic but a net magnetic moment is present along the two measured directions. However, data shows a clear FC drop in the measurement with **H** applied partially containing the [010] direction (Fig. 7(a)) that resemble a measurement of a pure antiferromagnet when **H** is along the antiferromagnetic axis. Although data supports the canted scenario sketched in Fig. 6(e), in the set of films discussed so far is not possible a conclusive measurement with **H** applied solely along the [010] direction due to the mixed contribution of two in-plane crystal domains. To circumvent this problem we prepared strained single domain films on STO(110) substrates. Details on the growth and structural characterization of similar films are described elsewhere [15]. In these films, (100)-textured, the [010] and [001] directions are contained in the plane of the sample. Measurements along [010] and [001] directions with **H** applied in the plane of the sample are plotted in Figs. 7(c) and 7(d), respectively. There is a perfect match between the ZFC and FC magnetic moment curves when **H** is applied along the [010] direction (Fig. 7(c)) indicating the absence of remanent net magnetic moment along the mentioned direction. In all, data in Fig. 7 clearly show that the net magnetic moment is only observed when the magnetic field is applied perpendicular to the spin-modulation, [010], direction.

The average departure of the magnetic moment away from the *b*-axis (canting angles) can be estimated from the measured remnant magnetization in the in-plane and out-of-plane configurations. Two magnetic loops corresponding to two (001)-oriented samples with different unit cell volumes are shown in the inset of Fig. 8. In this plot, the magnetic field is applied along the [110] direction. A larger remnant magnetic moment is found for the smaller unit cell supporting that the ferromagnetic response is increased by the unit cell contraction, also in agreement with our previous hypothesis based on the temperature dependence of the magnetization. From the values of the remnant magnetization and assuming that the total moment in the $Mn^{3+}$ atoms is 4 $\mu_B$, the component of magnetization along the [100] and [001] directions, perpendicular to the *b* axis, can be found by trigonometry. Notice the choice of polar ($\theta$) and azimuthal ($\phi$) angles definitions according to the sketch in Fig. 7e. Using the measurements with the magnetic field applied along the [110] direction, we estimate that the polar canting angle is higher than 1.2º in the most ferromagnetic sample. Data for the complete set of samples is shown in Fig. 8 where both the canting angle and the remnant magnetization are shown as a function of the unit cell volume. As shown in Fig. 8, naturally, this value gets reduced when films are progressively relaxed. For one sample we also show the calculated azimuthal canting angle by using the remnant magnetization of loops collected with magnetic field applied along the [001] direction. Data leads to a very similar value (empty symbol in Fig. 8) for the azimuthal canting angle. To conclude, magnetic anisotropy observed in the YMO films can be explained by a canting of the spins away from the [010] direction.

Before concluding we would like to stress that the observed anisotropic magnetic response, or more precisely, the observation of a net magnetization for a particular direction of the in-plane applied magnetic field, would not be expected if the finite remanent magnetization were produced by randomly distributed grain or domain boundaries.

## IV. SUMMARY and CONCLUSIONS

Using epitaxial strain and chemical pressure a substantially wide range of unit cell distortions in o-YMnO$_3$ has been achieved. We have shown that the observed net magnetic moment depends on the unit cell distortion involving both *b* and *c* distances which promote the occurrence of an uncompensated spin arrangement. Single domain fully textured YMO films have provided a unique opportunity to explore the magnetic anisotropy of the films, to demonstrate the intrinsic nature of the net magnetic moment and to show that it is related to spin canting out of the *b*-axis. More distortion implies larger ferromagnetic response; antiferromagnetism is recovered as lattice parameters tend to bulk. Therefore new, non-collinear magnetic structures have been stabilized by strain-driven distortions of the unit cell. Our findings provide a rationale to describe the observation of ferromagnetic response in o-YMnO$_3$ films, reported also in other orthorhombic manganites thin films. On these grounds, the strain-induced changes in the films would certainly resound on the dielectric properties as we have reported in an earlier work [14].

## Acknowledgements

Financial support by the Ministerio de Ciencia e Innovación of the Spanish Government [Projects MAT2008-06761-C03 and NANOSELECT CSD2007-00041] and Generalitat de Catalunya (2009 SGR 00376 ) are acknowledged.

Figure captions

FIG. 1. (a) θ/2θ scan for a YMCO sample (0.3 mbar, 58 nm). (b) φ-scans around the STO(111) and YMCO(111) reflections of the same sample. (c) Zoom of one YMCO(111) peak fitted by two Gaussian contributions.

FIG. 2. Reciprocal space maps around the o-YMCO(208) and (028) reflections for two pairs of samples (o-YMO and o-YMCO) of same thickness (~ 140 nm) but grown at (a) 0.3 and (b) 0.1 mbar.

FIG. 3. Reciprocal space maps around the YMO(208) and (028) reflections for (a) an as grown YMO sample (0.1 mbar, 30 nm) and after two subsequent annealings: (b) 12 hours at 800 ºC in air and (b) 2 hours at 700 ºC in 0.1 mbar Ar. Red lines denote the bulk positions for the reflections. Vertical dashed line is guide for the eye.

FIG. 4. FC magnetization curves (measured at 500 Oe) of (a) YMCO films, and (b) for an as-grown YMO sample (0.1 mbar, 30 nm) and after the first (air, 800 ºC) and second (Ar, 700 ºC) annealing mentioned in the text. In panel (a) ZFC branches are shown for two representative samples. Curves are labelled according to the unit cell volume of the samples.

FIG. 5. Magnetization (at 25 K and 500 Oe), measured for o-YMO and YMCO films versus lattice parameters (panels a and b) and the unit cell volume (panel c). Irrespective of the method used to tune the lattice parameters, the magnetization scales with unit cell contraction.

FIG. 6. Topographic images of (a) YMCO and (b) YMO films [both grown at 0.3 mbar, ≈ 100 nm] with partially superimposed pattern of grains in a different colour

scale. Panels (c) and (d) show the magnetization (25 K, 500 Oe) versus the grain density and surface roughness, respectively.

FIG. 7. ZFC-FC magnetic moment of YMO(001)/STO(001) films measured with H (500 Oe) applied (a) in-plane YMO[110] and (b) out-of-plane YMO[001] direction. ZFC-FC curves for single domain YMO(100)/STO(110) films along the (c) YMO[010] and (d) YMO[001] in-plane directions. Data show that net magnetization only occurs perpendicularly to [010] direction. Panel (e) shows a sketch of the spin-canted structure according to the experimental data. Notice the polar ($\theta$) and azimuthal ($\phi$) angles description in the sketch.

FIG. 8. Dependence of the remnant magnetization (right vertical axis) and calculated canting angles (left vertical axis) as a function of the unit cell volume. Data points represent the polar (solid circles) and azimuthal angles (empty circle) as defined in the text and sketch in Fig. 7. The data show that canting vanishes as the unit cell volume approaches the bulk value. Inset: a zoom of the magnetization loops of the two YMO samples with different unit cell volume (see legend), revealing the clear presence of a remanent magnetization.

Figure 1

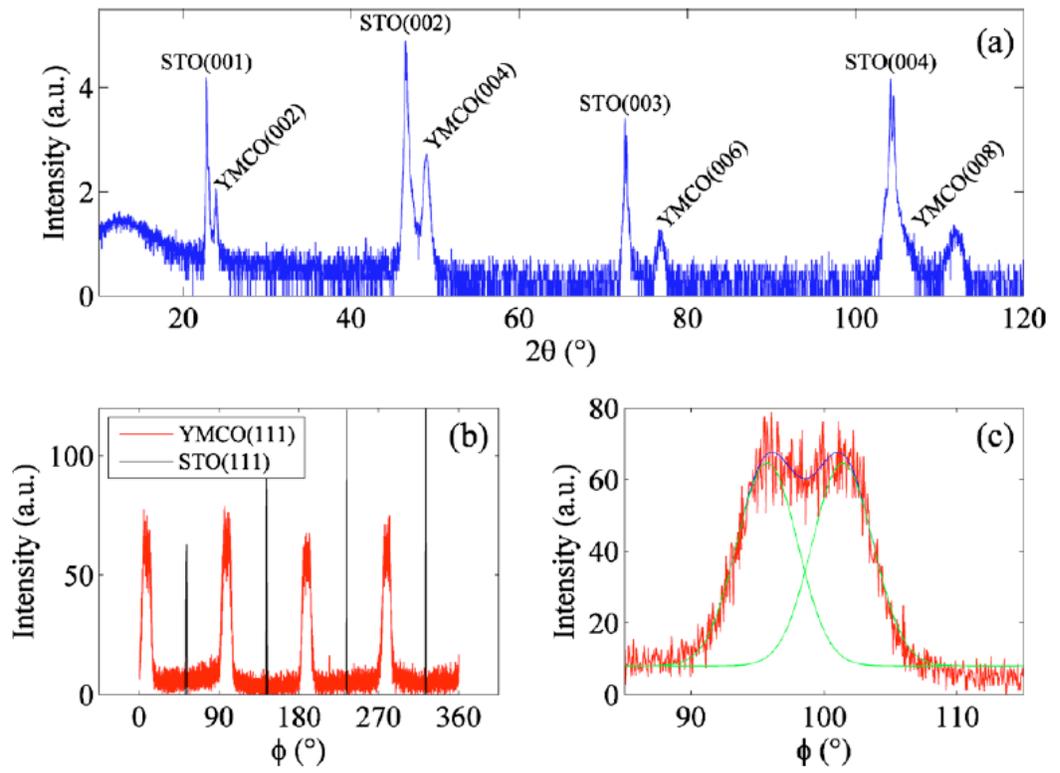

Figure 2

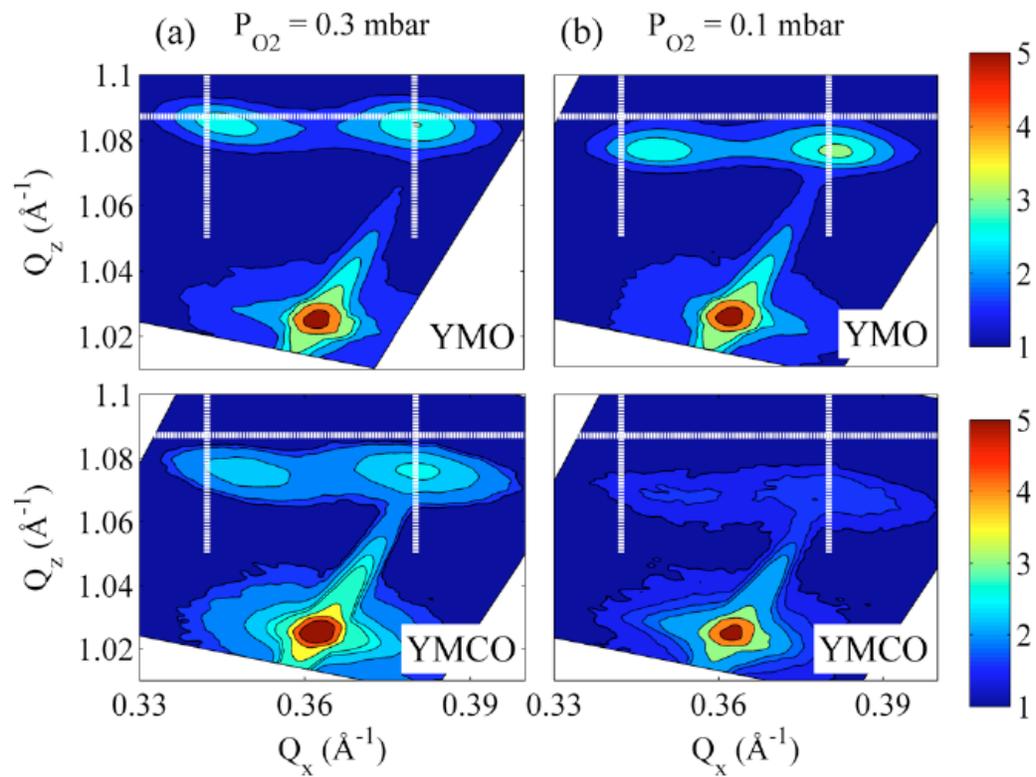

Figure 3

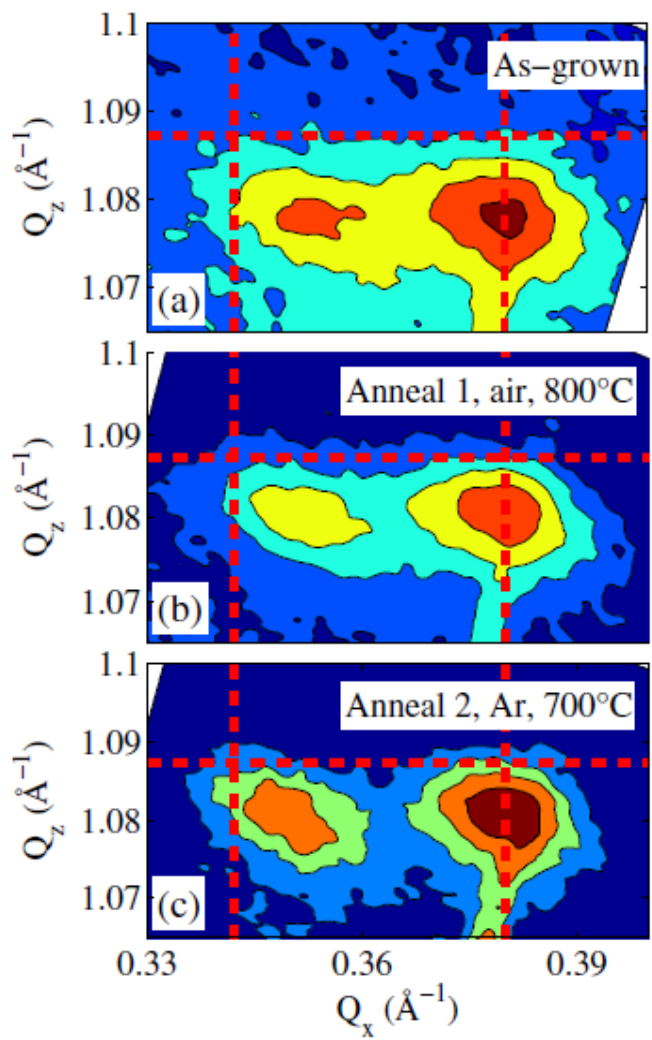

Figure 4

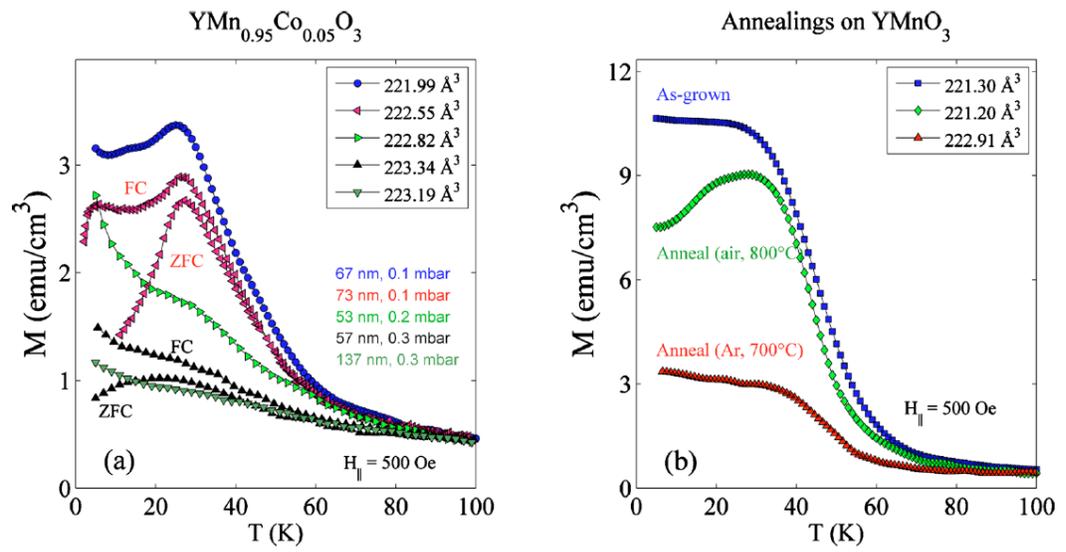

Figure 5

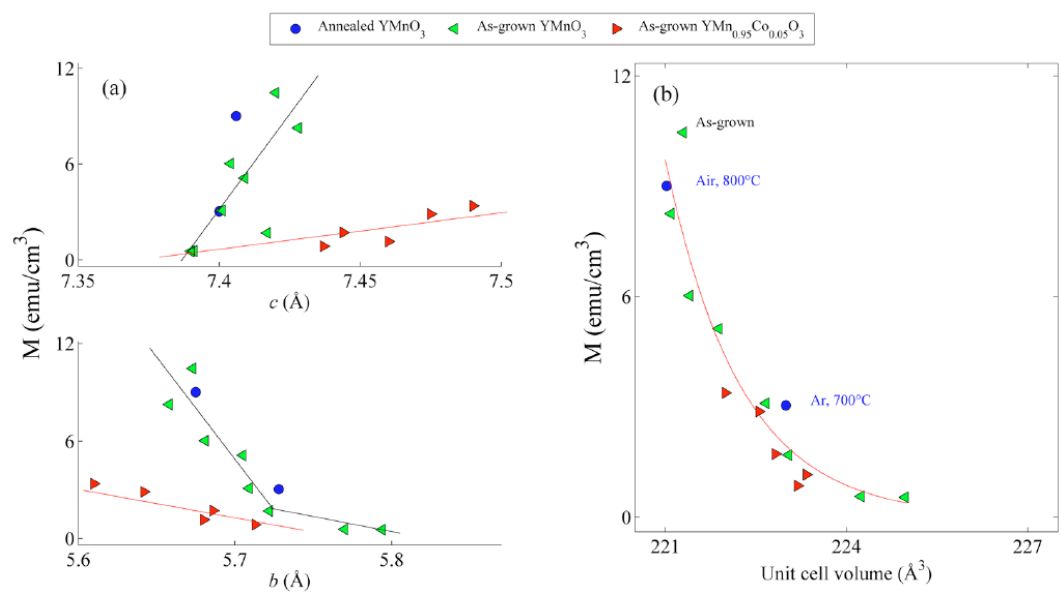

Figure 6

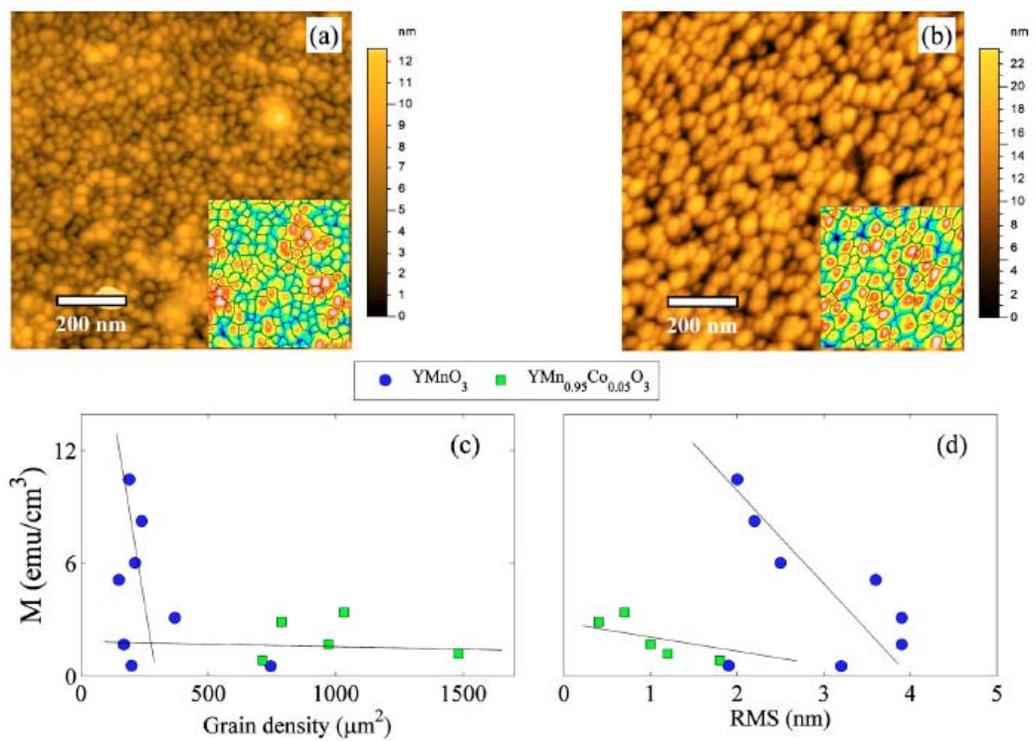

Figure 7

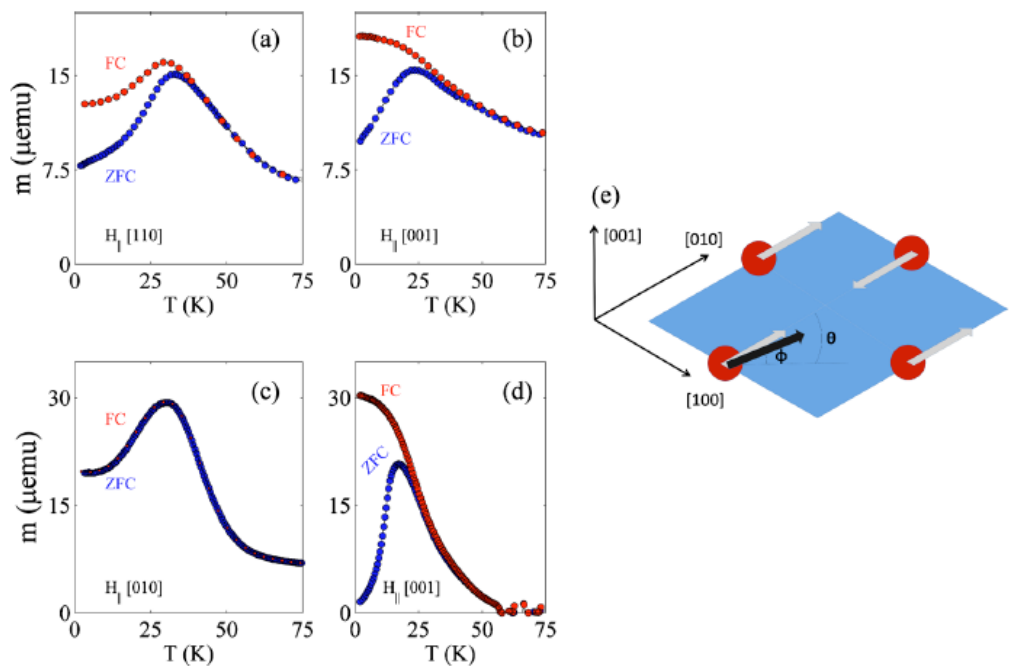

Figure 8

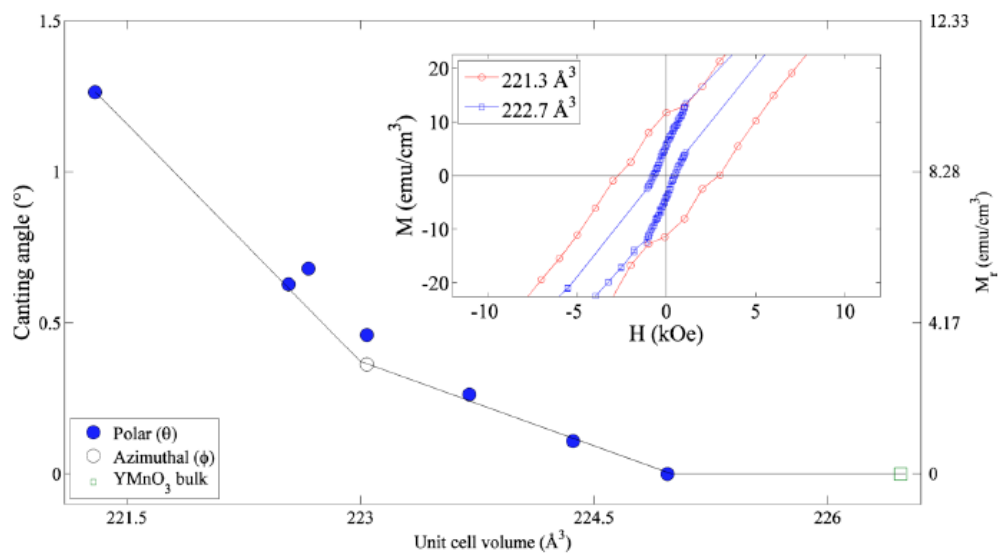